# Conservation of the spin and orbital angular momenta in electromagnetism

Konstantin Y. Bliokh[1,2], Justin Dressel[1,3], and Franco Nori[1,4]

[1]*CEMS, RIKEN, Wako-shi, Saitama 351-0198, Japan*
[2]*iTHES Research Group, RIKEN, Wako-shi, Saitama 351-0198, Japan*
[3]*Department of Electrical Engineering, University of California, Riverside, CA 92521, USA*
[4]*Physics Department, University of Michigan, Ann Arbor, Michigan 48109-1040, USA*

We review and re-examine the description and separation of the spin and orbital angular momenta (AM) of an electromagnetic field in free space. While the spin and orbital AM of light are not separately-meaningful physical quantities in orthodox quantum mechanics or classical field theory, these quantities are routinely measured and used for applications in optics. A meaningful quantum description of the spin and orbital AM of light was recently provided by several authors, which describes separately conserved and measurable *integral* values of these quantities. However, the electromagnetic field theory still lacks corresponding *locally*-conserved spin and orbital AM currents. In this paper, we construct these missing spin and orbital AM densities and *fluxes* that satisfy the proper continuity equations. We show that these are physically measurable and conserved quantities. These are, however, not Lorentz-covariant, so only make sense in the single laboratory reference frame of the measurement probe. The fluxes we derive improve the canonical (non-conserved) spin and orbital AM fluxes, and include a 'spin-orbit' term that describes the spin-orbit interaction effects observed in nonparaxial optical fields. We also consider both standard and dual-symmetric versions of the electromagnetic field theory. Applying the general theory to nonparaxial optical vortex beams validates our results and allows us to discriminate between earlier approaches to the problem. Our treatment yields the complete and consistent description of the spin and orbital AM of free Maxwell fields in both quantum-mechanical and field-theory approaches.

## 1. Introduction

It is known that light (electromagnetic waves or photons) can carry both spin and orbital angular momentum (AM) [1]. Locally, the spin density $\mathbf{S}$ is an intrinsic (i.e., origin-independent) quantity, which is associated with the local ellipticity of the polarization of light. In turn, the orbital AM density $\mathbf{L} = \mathbf{r} \times \mathbf{P}^O$ is a manifestly extrinsic (origin-dependent) and is produced by the corresponding canonical (orbital) momentum density $\mathbf{P}^O$. This momentum $\mathbf{P}^O$ is proportional to the phase gradient and can circulate in optical vortices [2–5]. Spin and orbital AM are widely used in classical and quantum optics as well-defined and separated degrees of freedom [1]. Optical experiments clearly show qualitatively different transfers of spin and orbital AM to small probe particles [6]. Namely, a small absorbing particle experiences a local *torque* proportional to $\mathbf{S}$ (that causes it to spin) and also a radiation-pressure *force* proportional to $\mathbf{P}^O$ (that causes it to orbit in optical vortices) [4,5,7,8]. Thus, spinning and orbital motions of a probe particle allow *operational measurements* of the separate spin and orbital AM densities in optical fields, see Fig. 1.

In theory, the separation of spin and orbital AM is unproblematic with paraxial monochromatic light, which is employed in most applications [1]. However, the self-consistent description and separation of the spin and orbital AM in generic electromagnetic fields is problematic and has caused a number of debates. Both the quantum mechanics of photons [9,10] and classical electromagnetic field theory [11] do not provide meaningful descriptions of the spin and orbital AM, but claim that only the total (spin+orbital) AM is a meaningful quantity. Indeed, the quantum-mechanical first-quantization operators of separated spin and orbital AM of light, $\hat{\mathbf{S}}$ and $\hat{\mathbf{L}} = \hat{\mathbf{r}} \times \hat{\mathbf{p}}$, are *inconsistent with the transversality condition* for photons, i.e., Maxwell's equations [9,10]. Furthermore, the spin and orbital parts of the conserved AM Noether current in electromagnetic field theory, $S^{\alpha\beta\gamma}$ and $L^{\alpha\beta\gamma} = r^\alpha T^{\beta\gamma} - r^\beta T^{\alpha\gamma}$ (where $T^{\alpha\beta}$ is the canonical stress-energy tensor), are *not conserved* separately [11]. In addition, these spin and orbital currents appear in canonical tensors that *cannot be made simultaneously gauge-invariant and Lorentz-covariant*.

Nonetheless, the local expectation values of the operators $\hat{\mathbf{S}}$ and $\hat{\mathbf{L}}$, as well as the pseudo-vectors $S_i = \frac{1}{2}\varepsilon_{ijk}S^{jk0}$ and $L_i = \frac{1}{2}\varepsilon_{ijk}L^{jk0}$ extracted from the spin and orbital AM tensors in the Coulomb gauge ($\varepsilon_{ijk}$ is the Levi-Civita symbol), yield the same values $\mathbf{S}$ and $\mathbf{L}$ that appear in optical experiments with monochromatic fields [12]. Moreover, the integral values of the spin and orbital AM, $\int \mathbf{S}\, dV$ and $\int \mathbf{L}\, dV$ (volume integrals for sufficiently localized fields are assumed), are conserved, i.e., time-independent, in free space [13]. This hints that the electromagnetic spin and orbital AM are separate physically-meaningful quantities, and that the fundamental problems with the quantum-mechanical and field-theory approaches can and should be overcome.

Indeed, the discrepancy between the quantum operators of the spin/orbital AM and the transversality of photons has been recently resolved [13–17]. It was shown, using both a second-quantization approach to $\mathbf{S}$ and $\mathbf{L}$ [13] and a first-quantization approach using $\hat{\mathbf{S}}$ and $\hat{\mathbf{L}}$ [14], that the suitably modified quantum-mechanical operators of the spin and orbital AM can be made consistent with both the field transversality and the measured expectation values. In the first-quantization formalism, the corrected spin and orbital AM operators acquire the form [14] $\hat{\tilde{\mathbf{S}}} = \hat{\mathbf{S}} - \hat{\boldsymbol{\Delta}}$ and $\hat{\tilde{\mathbf{L}}} = \hat{\mathbf{L}} + \hat{\boldsymbol{\Delta}}$, where $\hat{\boldsymbol{\Delta}}$ is a spin-orbit correction stemming from the transversality condition (similar corrected spin and orbital AM operators also appear for Dirac electron fields [18]).

This development is not yet a complete solution, however. The quantum-operator approach is based on the Fourier (momentum) representation and yields only *integral* expectation values of the spin and orbital AM. In contrast, the optical interaction with small particles or atoms requires a proper *local* description of the spin and orbital AM in terms of *densities* in real space. Furthermore, as the integral values $\int \mathbf{S}\, dV$ and $\int \mathbf{L}\, dV$ are conserved quantities, there should be a *continuity equation* describing the local transport and *fluxes* of the spin and orbital AM. Such continuity equation for optical spin $\mathbf{S}$ was discussed in several works [19–23], but most of these works have intrinsic discrepancies, and none of them derives the conserved spin and orbital AM currents as proper Noether AM currents within the electromagnetic field theory.

In this paper, we resolve this final fundamental problem in the description of the spin and orbital AM of light. Akin to the quantum-operator approach, we modify the separation of the spin and orbital parts of the canonical Noether AM current, $\tilde{S}^{\alpha\beta\gamma} = S^{\alpha\beta\gamma} - \Delta^{\alpha\beta\gamma}$ and $\tilde{L}^{\alpha\beta\gamma} = L^{\alpha\beta\gamma} + \Delta^{\alpha\beta\gamma}$, such that the modified tensors $\tilde{S}^{\alpha\beta\gamma}$ and $\tilde{L}^{\alpha\beta\gamma}$ satisfy a continuity equation and properly describe the spin and orbital AM densities $\mathbf{S}$ and $\mathbf{L}$. We show that this separation produces a meaningful local description of the spin and orbital AM densities and fluxes, and represents them as gauge-invariant (and, thus, observable) but *not* Lorentz-covariant quantities.

The latter fact is consistent with operational measurements, since a local probe particle will always single out the specific laboratory reference frame where it is at rest. Comparing our theory with other approaches and applying it to monochromatic optical fields validates our results and allows us to discriminate between various earlier attempts. Importantly, we find that the modification of the spin and orbital AM fluxes by the spin-orbit term $\Delta^{\alpha\beta\gamma}$ describes the spin-to-orbital AM conversion that is observed in nonparaxial optical fields [14,24] (see Fig. 2 below).

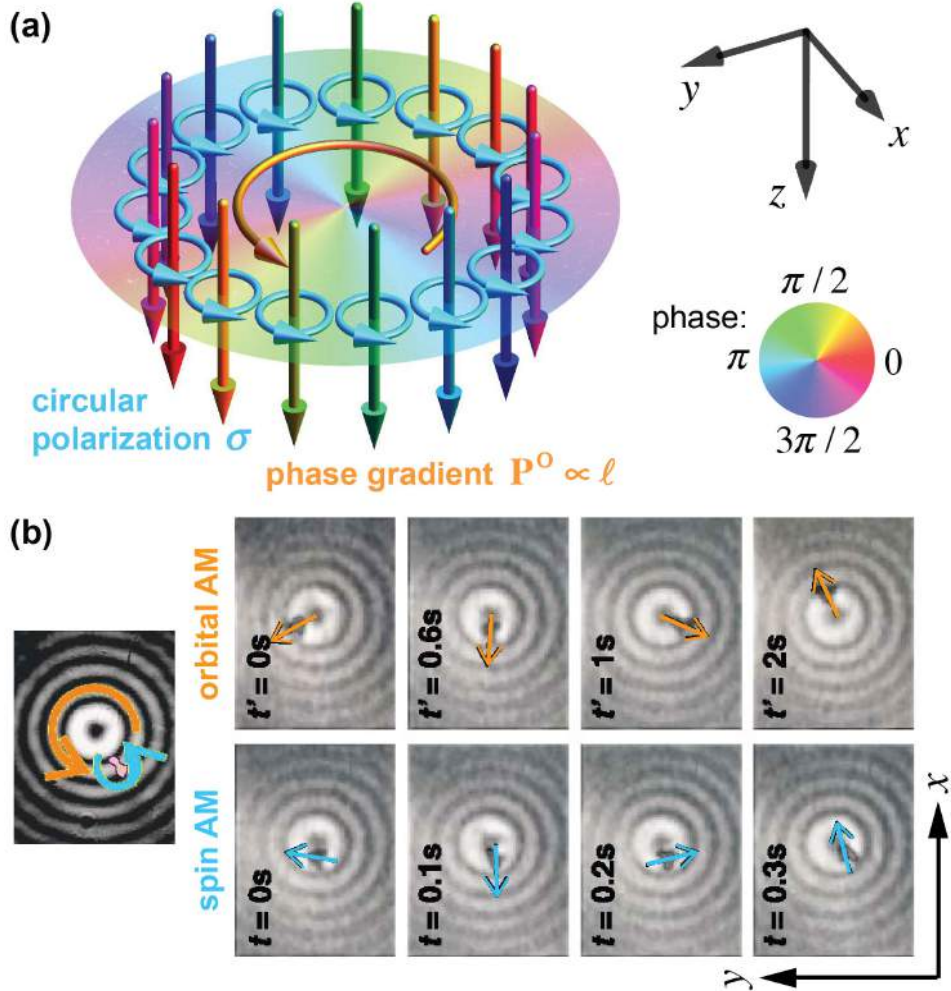

**Figure 1.** The spin and orbital AM densities of light are separately measurable quantities. The local ellipticity of polarization (times the normal to the polarization ellipse) determines the spin AM density $\bar{\mathbf{S}}$, Eq. (2.17). The local phase gradient of the field determines the canonical (orbital) momentum $\bar{\mathbf{P}}^O$, Eq. (2.18), and the corresponding orbital AM density $\mathbf{r} \times \bar{\mathbf{P}}^O$. A small probe dipole particle experiences both optical torque and radiation-pressure force, which are proportional to $\bar{\mathbf{S}}$ and $\bar{\mathbf{P}}^O$, respectively [4,5,7,8]. An example of the paraxial optical vortex beam with the left-hand circular polarization (ellipticity $\sigma = -1$) and charge-2 vortex ($\ell = -2$) generating the azimuthal phase gradient (orbital momentum) is shown in **(a)** (phase is color-coded). Experimental results **(b)** from [6c] demonstrate the spinning and orbital motion of a probe particle in such a paraxial vortex beam, which clearly indicate the separate local spin and orbital properties of the beam field.

The paper is organized as follows. In Section 2 we introduce the main equations and notations, and give an overview of the existing approaches to the problem, emphasizing their key shortcomings and subtle issues. We consider the conflict between the gauge invariance and Lorentz covariance, quantum-operator approaches, and the role of dual ('electric-magnetic') symmetry. In Section 3.1 we recall a general form of Noether conservation laws in electromagnetic field theory and indicate the way of constructing the spin and orbital AM conserved currents. Sections 3.2 and 3.3 show explicit calculations and results for these new conservation laws, using both the standard (electric-biased) and dual-symmetric electromagnetic theories. The latter one symmetrizes the electric and magnetic contributions [12,23,25]. In Section 4 we check our results by comparing them with other approaches and applying them to monochromatic optical fields (e.g., nonparaxial Bessel beams). Section 5 concludes the paper.

## 2. Overview of the problem

### *2.1. Basic notations and quantities.*

For the sake of simplicity, we use natural electrodynamical units $\varepsilon_0 = \mu_0 = c = 1$. Throughout the paper we assume Minkowski space-time $r^\alpha = (t, \mathbf{r})$ with metric tensor $g^{\alpha\beta} = \mathrm{diag}(-1,1,1,1)$. The Greek indices $\alpha, \beta, ...$ take on values 0,1,2,3, Latin indices $i, j, ...$ take on values 1,2,3, and summation over repeated indices is assumed. The four-dimensional and three-dimensional Levi-Civita symbols are $\varepsilon^{\alpha\beta\gamma\delta}$ and $\varepsilon_{ijk}$, and the Kronecker delta is $\delta_{ij}$.

The electric and magnetic fields are $\mathbf{E}(\mathbf{r},t)$ and $\mathbf{B}(\mathbf{r},t)$, and they satisfy free-space Maxwell equations:

$$\nabla \cdot \mathbf{E} = \nabla \cdot \mathbf{B} = 0, \quad \partial_t \mathbf{E} = \nabla \times \mathbf{B}, \quad \partial_t \mathbf{B} = -\nabla \times \mathbf{E}. \tag{2.1}$$

The first two equations (2.1) represent the *transversality condition*, i.e., the orthogonality of the Fourier-components of the fields to their $\mathbf{k}$-vectors.

Together with fields, we use the magnetic vector-potential $\mathbf{A}(\mathbf{r},t)$. In most cases we will assume the Coulomb gauge $\nabla \cdot \mathbf{A} = 0$, and the fields are expressed via this vector-potential as

$$\mathbf{E} = -\partial_t \mathbf{A}, \quad \mathbf{B} = \nabla \times \mathbf{A}. \tag{2.2}$$

Because of the *dual symmetry* between the electric and magnetic free-space Maxwell fields, we also use an *electric* vector-potential $\mathbf{C}(\mathbf{r},t)$, such that (assuming the Coulomb gauge $\nabla \cdot \mathbf{C} = 0$):

$$\mathbf{B} = -\partial_t \mathbf{C}, \quad \mathbf{E} = -\nabla \times \mathbf{C}. \tag{2.3}$$

Equations (2.2) and (2.3) show that magnetic and electric vector potentials are not independent quantities, but rather obey equations similar to Maxwell's equations (2.1) [12,23,25].

In covariant relativistic notation, the magnetic vector-potential becomes a part of the four-potential $A^\alpha = (A^0, \mathbf{A})$ [$A^\alpha = (0, \mathbf{A})$ in the Coulomb gauge], and the electromagnetic field is described by the anti-symmetric rank-2 tensor $F^{\alpha\beta} = \partial^\alpha \wedge A^\beta = (\mathbf{E}, \mathbf{B})$. The latter representation means that $F^{\alpha\beta} = -F^{\beta\alpha}$, $F^{0i} = E_i$, and $\frac{1}{2}\varepsilon_{ijk}F^{jk} = B_i$ is a pseudo-vector. There is also a *dual* field tensor $*F^{\alpha\beta} \equiv \frac{1}{2}\varepsilon^{\alpha\beta\gamma\delta}F_{\gamma\delta} = (\mathbf{B}, -\mathbf{E})$, which can be represented via the electric four-potential $C^\alpha = (C^0, \mathbf{C})$ [$C^\alpha = (0, \mathbf{C})$ in the Coulomb gauge] as $*F^{\alpha\beta} = \partial^\alpha \wedge C^\beta$. The covariant form of Maxwell's equations (2.1) is

$$\partial_\beta F^{\alpha\beta} = 0, \quad \partial_\beta *F^{\alpha\beta} = 0. \tag{2.4}$$

Considering monochromatic optical fields, we will use the complex field and vector-potential amplitudes $\mathbf{E}(\mathbf{r})$, $\mathbf{A}(\mathbf{r})$, etc., defined as

$$\mathbf{E}(\mathbf{r},t) = \operatorname{Re}\left[\mathbf{E}(\mathbf{r})e^{-i\omega t}\right], \quad \mathbf{A}(\mathbf{r},t) = \operatorname{Re}\left[\mathbf{A}(\mathbf{r})e^{-i\omega t}\right], \quad \text{etc.} \tag{2.5}$$

In this case, the Coulomb-gauge vector-potentials have simple relations to the fields:

$$\mathbf{A}(\mathbf{r}) = -i\omega^{-1}\mathbf{E}(\mathbf{r}), \quad \mathbf{C}(\mathbf{r}) = -i\omega^{-1}\mathbf{B}(\mathbf{r}). \tag{2.6}$$

As usual in optics, the bilinear quantities calculated for monochromatic fields will be averaged over oscillations in time.

### 2.2. Spin and orbital AM densities. Gauge invariance versus Lorentz covariance.

Classical electromagnetic field theory produces a manifestly covariant canonical (Noether) rank-3 tensor of the AM current [11], $M^{\alpha\beta\gamma} = S^{\alpha\beta\gamma} + r^{\alpha}T^{\beta\gamma} - r^{\beta}T^{\alpha\gamma} \equiv S^{\alpha\beta\gamma} + L^{\alpha\beta\gamma}$ that represents the sum of the spin and orbital contributions, $S^{\alpha\beta\gamma}$ and $L^{\alpha\beta\gamma}$. The spatial densities of the spin and orbital AM are given by the pseudo-vectors $S_i = \frac{1}{2}\varepsilon_{ijk}S^{jk0}$ and $L_i = \frac{1}{2}\varepsilon_{ijk}L^{jk0}$. Explicitly, these have the form

$$\mathbf{S} = \mathbf{E}\times\mathbf{A}, \quad \mathbf{L} = \mathbf{r}\times\mathbf{P}^O = \mathbf{E}\cdot(\mathbf{r}\times\nabla)\mathbf{A}, \tag{2.7}$$

where $\mathbf{P}^O = \mathbf{E}\cdot(\nabla)\mathbf{A}$ is the canonical (orbital) momentum density of the field [2–5,12], and we adopt the notation $\mathbf{X}\cdot(\mathbf{Z})\mathbf{Y} = X_i Z Y_i$ for any quantities $\mathbf{X}$, $\mathbf{Y}$, $\mathbf{Z}$.

The quantities (2.7) are gauge-dependent (and, hence, non-observable) since they explicitly involve the vector-potential $\mathbf{A}$, i.e., the spatial part of the four-potential $A^{\alpha} = (A^0, \mathbf{A})$. To provide gauge-invariant quantities, modified definitions of the spin and orbital densities are typically used [10]:

$$\mathbf{S} = \mathbf{E}\times\mathbf{A}_{\perp}, \quad \mathbf{L} = \mathbf{r}\times\mathbf{P}^O = \mathbf{E}\cdot(\mathbf{r}\times\nabla)\mathbf{A}_{\perp}. \tag{2.8}$$

Here the vector-potential is represented as a sum $\mathbf{A} = \mathbf{A}_{\perp} + \mathbf{A}_{\parallel}$ of the 'transverse' and 'longitudinal' parts, which obey the conditions $\nabla\cdot\mathbf{A}_{\perp} = 0$ and $\nabla\times\mathbf{A}_{\parallel} = 0$, respectively. Since gauge transformations of the vector-potential involve only the longitudinal part $\mathbf{A}_{\parallel}$ and the time component $A^0$, equations (2.8) are gauge-invariant. The definitions (2.8) coincide with (2.7) if one sets the *Coulomb gauge* in a given frame:

$$A^{\alpha} = (0, \mathbf{A}) \equiv (0, \mathbf{A}_{\perp}), \quad \nabla\cdot\mathbf{A} = 0. \tag{2.9}$$

Importantly, the gauge-invariant definition (2.8) breaks the Lorentz covariance of the original quantities (2.7) originating from the covariant tensor currents $S^{\alpha\beta\gamma}$ and $L^{\alpha\beta\gamma}$. Indeed, the transverse part of the vector potential, $\mathbf{A}_{\perp}(\mathbf{r},t)$, is not transformed covariantly and, when given in one reference frame, it becomes essentially *nonlocal* in another reference frame [10,16]. Nonetheless, the *integral* values of the spin and orbital AM, as defined via (2.8), are well-defined quantities, which can be calculated in any reference frame. Moreover, for free-space Maxwell fields, these are *conserved* quantities [13]:

$$\partial_t \int \mathbf{S}\,dV = 0, \quad \partial_t \int \mathbf{L}\,dV = 0. \tag{2.10}$$

Thus, adopting the Coulomb gauge (2.9), the canonical field-theory tensors yield meaningful spin and orbital AM of the electromagnetic field. These are *not* Lorentz-covariant and can be introduced only in one chosen reference frame. However, appealing to the optical experience and applications, does one really need Lorentz covariance for local densities of dynamical properties of the field? The operational measurements of the spin, orbital and other densities via local probes [4–8] always *single out one particular laboratory reference frame*, where the probe is at rest. Therefore, to compare theory with experiment, one must properly define these densities only in one *laboratory* reference frame. In addition, most optical applications deal with *monochromatic* electromagnetic waves, which also depend on a chosen laboratory reference frame (a transverse Lorentz boost makes optical beams non-monochromatic [26]). Based on these considerations, *in what follows we use general covariant notations of the field theory, but adopt the Coulomb gauge condition (2.9) in all explicit calculations* to isolate the gauge-invariant transverse part of the vector-potential. This means that our theory is actually gauge-invariant ($\mathbf{A}$ implies $\mathbf{A}_\perp$) but *not* Lorentz-covariant, i.e., it makes sense only in the laboratory reference frame of a local probe.

It is important to note that in this paper, following previous works [12,15,23,25], we imply 'locality' of the spin and orbital AM densities in terms of the transverse vector potentials. Although the latters are themselves integral functions of the field strengths [10,16], this nonlocal relationship does not affect our conservation laws that treat the transverse vector potentials as fundamental. Furthermore, in the most practically important case of monochromatic optical fields, the transverse vector potentials become *locally* related to the field strengths, as shown in Eqs. (2.6).

### 2.3. Quantum approaches.

It is instructive to review the quantum-operator approach to the spin and orbital AM of free electromagnetic fields. There are two levels of quantum formalism: (i) the first quantization, which deals with operators of dynamical variables acting on the *classical* electromagnetic field (i.e., this is essentially a representation of classical electrodynamics) and (ii) the second quantization, which quantizes the fields and make them quantum operators acting on the Fock states of photons.

In the first-quantization approach, the AM operator underlying the spin and orbital AM (2.8) is [9,10]

$$\hat{\mathbf{M}} = \hat{\mathbf{S}} + \hat{\mathbf{r}} \times \hat{\mathbf{p}} \equiv \hat{\mathbf{S}} + \hat{\mathbf{L}}. \tag{2.11}$$

Here $\hat{\mathbf{S}}$ is the spin-1 operator given by the $3 \times 3$ matrix generators of $SO(3)$ rotations, whereas $\hat{\mathbf{r}}$ and $\hat{\mathbf{p}}$ are the canonical coordinate and momentum operators. The operators $\hat{\mathbf{S}}$ and $\hat{\mathbf{L}}$ obey the standard $so(3)$ rotation algebra. However, when acting on free electromagnetic fields they do not preserve their transversality, i.e.,

$$\nabla \cdot \left( \hat{S}_i \mathbf{E} \right) = -\nabla \cdot \left( \hat{L}_i \mathbf{E} \right) \neq 0. \tag{2.12}$$

The reason for this is that $\hat{\mathbf{S}}$ generates rotations of only directions of the field vectors (but not their spatial distributions), while $\hat{\mathbf{L}}$ rotates only the spatial distribution of the field (but not its direction) [10,15]. Therefore canonical spin and orbital AM operators are not consistent with Maxwell's equations (2.1). In contrast, the total AM operator $\hat{\mathbf{M}}$ generates rotations of the whole field $\mathbf{E}(\mathbf{r},t)$ (both directions and distributions), and is consistent with the transversality. Due to this, most textbooks in quantum electrodynamics claim that the spin and orbital parts of the AM of light are not separately meaningful, and only the total AM of a photon makes sense [9,10].

In 1994, van Enk and Nienhuis [13] showed that, despite the fundamental problems with the operators $\hat{\mathbf{S}}$ and $\hat{\mathbf{L}}$, the second quantization of the integral spin and orbital AM, $\int \mathbf{S} \, dV$ and $\int \mathbf{L}\,dV$, based on the densities (2.8) results in meaningful field operators of the spin and orbital AM $\hat{\mathcal{S}}$ and $\hat{\mathcal{L}}$. These second-quantization operators are fully consistent with the transversality and Maxwell equations, but have unusual commutation relations that are different from the $so(3)$ algebra [13]. Considering a local quantum dipole interaction of light with an atom, van Enk and Nienhuis found that "both spin and orbital AM of a photon are well defined and separately measurable".

In 2010, Bliokh et al. [14] found a consistent first-quantized description of the spin and orbital AM of light (similar theory for Dirac electron fields is described in [18]). The problem with $\hat{\mathbf{S}}$ and $\hat{\mathbf{L}}$ was resolved by modifying the separation (2.11) into spin and orbital parts to make them consistent with the field transversality. This modification can be interpreted as a 'projection' of the canonical operators onto the transversality subspace. The resulting modified spin and orbital AM operators acquire the form:

$$\hat{\tilde{\mathbf{S}}} = \hat{\mathbf{S}} - \hat{\mathbf{\Delta}} = \hat{\boldsymbol{\kappa}}\hat{H}, \quad \hat{\tilde{\mathbf{L}}} = \hat{\mathbf{L}} + \hat{\mathbf{\Delta}} = \hat{\mathbf{r}}' \times \hat{\mathbf{p}}. \qquad (2.13)$$

Here $\hat{\mathbf{\Delta}} = -\hat{\boldsymbol{\kappa}} \times (\hat{\boldsymbol{\kappa}} \times \hat{\mathbf{S}})$ is the spin-orbit correction term, $\boldsymbol{\kappa} = \hat{\mathbf{p}}/\hat{p}$ ($\hat{p}$ is the scalar total momentum operator), $\hat{H} = \hat{\boldsymbol{\kappa}} \cdot \hat{\mathbf{S}}$ is the *helicity* operator, and $\hat{\mathbf{r}}' = \hat{\mathbf{r}} + (\hat{\mathbf{p}} \times \hat{\mathbf{S}})/\hat{p}^2$ is the so-called Pryce position operator for photons [27], well-known in the theory of relativistic spinning particles [28]. The modified operators (2.13) are properly consistent with the field transversality and obey the same non-canonical commutation relations as the second-quantization operators AM $\hat{\mathcal{S}}$ and $\hat{\mathcal{L}}$ in [13]. Remarkably, the modified operators (2.13) produce the *same* expectation values of the spin and orbital AM as the canonical operators $\hat{\mathbf{S}}$ and $\hat{\mathbf{L}}$, so this modification does not affect any observable quantities, which are still based on Eqs. (2.8). At the same time, the operators $\hat{\tilde{\mathbf{S}}}$ and $\hat{\tilde{\mathbf{L}}}$ acquire a particularly simple diagonal form in the helicity (momentum-space) representation. They facilitate Fourier-space calculations and illuminate spin-orbit conversion processes originating from Berry-phase effects [14] (see subsection 4.2 and Fig. 2 below).

Other aspects of the spin and orbital AM of light in quantum formalisms were also analysed by Barnett [15], Bialynicki-Birula [16], and Fernandez-Corbaton et al. [17]. These works agree with the approaches of [13] and [14] described above. Thus, the problem with the quantum-operator description of the integral spin and orbital AM of light seems to be resolved.

It should be emphasized, however, that the operators (2.13) allow efficient calculations of the *integral* values of the spin and orbital AM, but not their local values in the generic case. Indeed, these operators can be easily used in the momentum (Fourier) representation [14]. In the coordinate (real-space) representation, they become *nonlocal* because of the $1/\hat{p}$ operator, and do not yield densities of the spin and orbital AM. An important exception is the case of *monochromatic* fields (2.5). In this case, the complex field and vector-potential amplitudes become proportional to each other with a factor of frequency $\omega$, Eq. (2.6). Then, the time-averaged spin and orbital AM densities (2.8) acquire the form of local expectation values of operators $\hat{\mathbf{S}}$ and $\hat{\mathbf{L}}$ or $\hat{\tilde{\mathbf{S}}}$ and $\hat{\tilde{\mathbf{L}}}$ with the 'wave function' $\psi \propto \omega^{-1/2}\mathbf{E}(\mathbf{r})$ [5,8,12]. For a generic non-monochromatic field the operation $1/\omega \propto 1/\hat{p}$ is nonlocal [10,16].

### 2.4. Spin and orbital AM currents in field theory.

The consistent quantum formalism does not fully resolve the problem of obtaining a *local* description of the spin and orbital AM in electromagnetism. To address this remaining problem, we must return to a field theory description. The canonical spin and orbital AM tensors $S^{\alpha\beta\gamma}$ and $L^{\alpha\beta\gamma}$ in the Coulomb gauge (2.9) yield the spin and orbital AM densities (2.8), which correspond to optical experiments and conserved integral values (2.10), consistent with the quantum-mechanical approach. However, the canonical tensors $S^{\alpha\beta\gamma}$ and $L^{\alpha\beta\gamma}$ do not yield proper *fluxes* of the spin and orbital AM – they do not satisfy the continuity equation, i.e., the local conservation law for the spin and orbital AM:

$$\partial_\gamma S^{\alpha\beta\gamma} = -\partial_\gamma L^{\alpha\beta\gamma} \neq 0. \tag{2.14}$$

This problem resembles the inconsistency of the canonical operators $\hat{\mathbf{S}}$ and $\hat{\mathbf{L}}$, Eq. (2.11), with the transversality, Eq. (2.12). Equation (2.14) shows that the canonical spin and orbital AM currents are not conserved separately, while their sum is: $\partial_\gamma M^{\alpha\beta\gamma} = 0$. Non-conserved currents (2.14) conflict with the conservation of the integral values (2.10); one could expect that there should be proper continuity equations that describe the local transport of the separately-conserved spin and orbital AM.

There have been several attempts to suggest such a continuity equation for optical spin. In 2001, Alexeyev *et al.* [19] suggested a continuity equation for spin using complex Maxwell fields, with similar equations later considered by others [21,22]. However, in these papers, the authors considered *complex* time-dependent fields in Maxwell equations, while the proper fields must be real. This is an intrinsically inconsistent approach, which could make sense only in the most trivial case of monochromatic fields with time-independent complex amplitudes. In 2002, Barnett [20] considered the flux of the total AM from $M^{\alpha\beta\gamma}$ and suggested the separation "by hand" of the spin and orbital parts. However, the continuity equations for the spin and orbital parts were not provided. Furthermore, the ratios of spin and orbital fluxes to the energy flux, calculated in [20] for non-paraxial optical beams, contradict calculations of spin and orbital AM in non-paraxial Bessel beams in [14]. Namely, the spin-to-orbital conversion in non-paraxial fields is absent in [20], while it is clearly observed in experiments and described in theory [24,14]. Finally, in 2012 Cameron *et al.* [23] derived the continuity equation for the electromagnetic spin using an extension of the local *helicity* conservation law. However, the helicity conservation follows from the so-called *dual* symmetry between electric and magnetic fields [12,23,25,29], and not from Poincaré symmetries of space-time. Therefore, the continuity equation derived in [23] works only for the dual-symmetrized spin but not for the standard spin density (2.7) and (2.8) (see the next subsection). Thus, the spin AM continuity equation has never appeared as a proper conservation law following from the field-theory AM tensor. Furthermore, the continuity equation for the orbital AM has never been considered at all.

### *2.5. The role of the dual 'electric-magnetic' symmetry.*

So far, we considered the AM problem using only the *electric* field and vector-potential $\mathbf{A}$. However, free-space Maxwell electromagnetism possesses an important symmetry between the electric and magnetic properties. This is the so-called *dual* symmetry [12,23,25,29]. To take this symmetry into account, one has to consider electric and magnetic fields, $\mathbf{E}$ and $\mathbf{B}$, on equal footing. This naturally involves two vector-potentials, $\mathbf{A}$ and $\mathbf{C}$, Eqs. (2.2) and (2.3). Recently, we showed [12] (see also [25]) that choosing a suitable Lagrangian, one can construct the dual-symmetric free-space electromagnetic field theory, preserving the symmetry between electric and magnetic properties.

The discrete form of the dual transformation reads

$$\mathbf{E} \to \mathbf{B}, \quad \mathbf{B} \to -\mathbf{E},$$
$$\mathbf{A} \to \mathbf{C}, \quad \mathbf{C} \to -\mathbf{A}. \tag{2.15}$$

The dynamical characteristics of the electromagnetic field become symmetrized with respect to the transformation (2.15) in the dual-symmetric electromagnetism [2,12,15,23,25,29]. In particular, spin and orbital AM densities (2.8) become

$$\mathbf{S} = \frac{1}{2}(\mathbf{E} \times \mathbf{A} + \mathbf{B} \times \mathbf{C}), \quad \mathbf{L} = \mathbf{r} \times \mathbf{P}^{\mathrm{O}} = \frac{1}{2}\big[\mathbf{E} \cdot (\mathbf{r} \times \nabla)\mathbf{A} + \mathbf{B} \cdot (\mathbf{r} \times \nabla)\mathbf{C}\big]. \tag{2.16}$$

Such dual symmetrization does not change the integral values of the spin and orbital AM, $\int \mathbf{S}\, dV$ and $\int \mathbf{L}\, dV$ [15,30], but their densities become different. For example, considering the time-averaged values of the spin density and orbital momentum density in a monochromatic field (2.5) and (2.6), the standard (electric-biased) and dual-symmetric versions of electromagnetism yield the following quantities [12]:

$$\overline{\mathbf{S}}^{\text{standard}} = \frac{1}{2\omega}\operatorname{Im}\big(\mathbf{E}^* \times \mathbf{E}\big), \quad \overline{\mathbf{S}}^{\text{dual}} = \frac{1}{4\omega}\operatorname{Im}\big(\mathbf{E}^* \times \mathbf{E} + \mathbf{B}^* \times \mathbf{B}\big). \tag{2.17}$$

$$\overline{\mathbf{P}}^{\mathrm{O\,standard}} = \frac{1}{2\omega}\operatorname{Im}\big[\mathbf{E}^* \cdot (\nabla)\mathbf{E}\big], \quad \overline{\mathbf{P}}^{\mathrm{O\,dual}} = \frac{1}{4\omega}\operatorname{Im}\big[\mathbf{E}^* \cdot (\nabla)\mathbf{E} + \mathbf{B}^* \cdot (\nabla)\mathbf{B}\big]. \tag{2.18}$$

The standard and dual-symmetric densities in Eqs. (2.17) or (2.18) are equivalent for paraxial propagating fields (as, e.g., in Fig. 1), but can be significantly different in nonparaxial or other complex fields [2]. For instance, the two definitions (2.17) result, respectively, in zero and nonzero transverse spin in evanescent TE waves [31,5]. In what follows, we will use the "standard" and "dual" superscript only when needed to emphasize the difference between the two theories.

The question whether one should use the dual-symmetric versions of these quantities is another subtle issue. On the one hand, the fundamental dual symmetry of free-space Maxwell theory makes the dual-symmetric definitions more natural [2,15] and self-consistent [12]. For instance, the standard definition of spin in (2.17) implies that a rotating electric field produces spin AM, while a rotating magnetic field does not. This would be bizarre.

On the other hand, if we rely on *experimental* measurements of the spin density, we should consider the interaction of the electromagnetic field with a small probe particle or other measuring device (see Fig. 1). Importantly, any measuring device represents *matter*, and matter is *not* dual-symmetric in electromagnetism. There are *electric* charges but no magnetic charges. Therefore, a typical point-dipole particle or an atom is coupled to the *electric* rather than magnetic field (see, e.g., [2,5,12,13]). Accordingly, such an electric-dipole probe will measure the electric part of the spin, i.e., its standard definition. Similarly, a magnetic-dipole particle or another particle with complex properties would 'measure' magnetic spin density or more sophisticated quantities [8].

Thus, the fundamental dual symmetry and structure of free-space fields implies dual-symmetric definitions of all meaningful quantities, while practical applications might require alternative quantities, which depend on the character of light-matter interaction. The electric-dipole interaction involves the standard 'electric' spin. (See also discussion in [12].)

In this paper we consider both standard ('electric-biased') and dual-symmetric versions of electromagnetism. Since all calculations are quite similar in the two theories [12], we perform explicit calculations for the standard theory, and then show the final results of the dual-symmetric calculations. As we will see, locally conserved spin and orbital AM currents can be equally constructed within *both* approaches, i.e., independently of the dual symmetry.

## 3. Conserved spin and orbital AM currents

### 3.1. General covariant form.

We are now in a position to construct the proper field-theory description of the electromagnetic spin and orbital AM currents. We start with the main local conservation laws in electromagnetic field theory [11,12]. In this subsection we present these in a general tensor form, without explicit expressions in terms of fields.

As is well known, Noether's theorem results in the conservation laws associated with continuous symmetries of the Lagrangian or equations of motion. First, the symmetry with respect to translations in space-time result in momentum-energy conservation. Applying Noether's theorem yields the *canonical* stress-energy tensor $T^{\alpha\beta}$ and the corresponding conservation law:

$$\partial_\beta T^{\alpha\beta} = 0, \quad T^{\alpha\beta} \neq T^{\beta\alpha}. \tag{3.1}$$

Note that the canonical stress-energy tensor is non-symmetric. This tensor contains the four-vector $P^\alpha = T^{0\alpha} = (W, \mathbf{P}^O)$ representing the canonical four-momentum density, including the energy density $W = (\mathbf{E}^2 + \mathbf{B}^2)/2$ and the orbital momentum density $\mathbf{P}^O$.

Second, the symmetry with respect to rotations of the Minkowski space-time generates the relativistic angular-momentum conservation. It is described by the rank-3 $\alpha\beta$-antisymmetric AM tensor $M^{\alpha\beta\gamma}$:

$$\partial_\gamma M^{\alpha\beta\gamma} = 0, \quad M^{\alpha\beta\gamma} = -M^{\beta\alpha\gamma}. \tag{3.2}$$

The AM tensor (3.2) is related to the stress-energy tensor (3.1) as

$$M^{\alpha\beta\gamma} = r^\alpha T^{\beta\gamma} - r^\beta T^{\alpha\gamma} + S^{\alpha\beta\gamma} \equiv L^{\alpha\beta\gamma} + S^{\alpha\beta\gamma}, \tag{3.3}$$

where $S^{\alpha\beta\gamma}$ is the so-called *spin tensor*. The form of Eq. (3.3) suggests that the AM tensor consists of an orbital (extrinsic) part $L^{\alpha\beta\gamma}$ and a spin (intrinsic) part $S^{\alpha\beta\gamma}$. However, these two parts are not conserved separately. Indeed, substituting Eq. (3.3) into (3.2) and using Eq. (3.1), we obtain

$$\partial_\gamma S^{\alpha\beta\gamma} = -\partial_\gamma L^{\alpha\beta\gamma} = T^{\alpha\beta} - T^{\beta\alpha} \neq 0. \tag{3.4}$$

In 1939, Belinfante [32] suggested a useful procedure to symmetrize the canonical stress-energy tensor by adding a suitable total divergence to the canonical stress-energy tensor. This procedure results in a *symmetric* stress-energy tensor $\mathcal{T}^{\alpha\beta}$ and conservation law:

$$\mathcal{T}^{\alpha\beta} = T^{\alpha\beta} + \partial_\gamma K^{\alpha\beta\gamma}, \quad \partial_\beta \mathcal{T}^{\alpha\beta} = 0, \quad \mathcal{T}^{\alpha\beta} = \mathcal{T}^{\beta\alpha}, \tag{3.5}$$

where the tensor $K^{\alpha\beta\gamma}$ is constructed from the spin tensor (3.3):

$$K^{\alpha\beta\gamma} = \frac{1}{2}\left(S^{\beta\gamma\alpha} + S^{\alpha\gamma\beta} - S^{\alpha\beta\gamma}\right). \tag{3.6}$$

The symmetrized tensor (3.5) contains the four-momentum density $\mathcal{P}^\alpha = \mathcal{T}^{0\alpha} = (W, \boldsymbol{\mathcal{P}})$ including the Poynting vector $\boldsymbol{\mathcal{P}} = \mathbf{E} \times \mathbf{B}$ [11]. The corresponding *symmetrized* AM tensor can then be constructed from the symmetric stress-energy tensor (3.5):

$$\mathcal{M}^{\alpha\beta\gamma} = r^\alpha \mathcal{T}^{\beta\gamma} - r^\beta \mathcal{T}^{\alpha\gamma}, \quad \partial_\beta \mathcal{M}^{\alpha\beta\gamma} = 0. \tag{3.7}$$

It might seem that this AM tensor contains only the orbital part, but actually it also includes the spin because the integral values of the AM (3.3) and (3.7) coincide for sufficiently localized fields: $\int M^{\alpha\beta 0} dV = \int \mathcal{M}^{\alpha\beta 0} dV = \text{const}$. Nonetheless, separating the spin and orbital parts of the AM is problematic with the symmetrized tensor (3.7). Furthermore, we emphasize that although the symmetrized stress-energy tensor (3.5) is typically considered in field theory as physically meaningful (the source of the gravitational field), it is the canonical momentum density $\mathbf{P}^\mathrm{O}$ stemming from the *canonical* tensor (3.1) that appears in optical and quantum-mechanical measurements of the momentum density of light. See, e.g., discussions in [4,5,8,12], quantum weak measurements in [33], and the transfer of optical 'super-momentum' $|\mathbf{P}^\mathrm{O}|/W > 1$ (impossible with the Poynting vector, $|\mathbf{\mathcal{P}}|/W \leq 1$) in [34].

In addition to the conservation laws associated with Poincaré symmetries (i.e., transformations of the space-time), there is one more fundamental conservation law for free-space Maxwell fields. Namely, there is conservation of the *helicity*, associated with the continuous version of the internal dual symmetry between the electric and magnetic parts of the free field [12,23,25,29]. The conservation of the helicity current can be written as

$$\partial_\alpha J^\alpha = 0, \quad J^\alpha = \left(H, \mathbf{S}^\mathrm{dual}\right). \tag{3.8}$$

Here $J^\alpha$ is a four-pseudovector, with its zero component $H$ being the helicity density pseudo-scalar and $\mathbf{S}^\mathrm{dual}$ being the helicity flux pseudo-vector.

Importantly, for the dual-symmetric formulation of electromagnetism [12,23,25], this helicity flux precisely coincides with the pseudo-vector of the spin density obtained from the spin tensor: $S_i^\mathrm{dual} = \frac{1}{2}\varepsilon_{ijk}S^{jk0\,\mathrm{dual}}$. Thus, the same *dual-symmetric* spin density can be obtained either from the AM tensor or from the helicity current.

Recently, Cameron *et al.* [23] suggested an extension of the conserved helicity four-current (3.8) to a rank-3 pseudo-tensor similar to the so-called Lipkin's zilch pseudo-tensor [35]:

$$\partial_\gamma J^{\alpha\beta\gamma} = 0, \quad J^{\alpha\beta\gamma} = J^{\beta\alpha\gamma}. \tag{3.9}$$

Here $J^{00\alpha} = J^\alpha$, and equation (3.9) also includes a continuity equation for the spin $\mathbf{S}^\mathrm{dual}$:

$$\partial_t \mathbf{S}^\mathrm{dual} + \partial_j \Sigma_{ij}^\mathrm{dual} = 0, \tag{3.10}$$

where $\Sigma_{ij}^\mathrm{dual} = J^{0ij}$. Thus, the dual-symmetric spin (following from the helicity flux) satisfies the continuity equation (3.10) despite the non-conserved spin tensor (3.4). However, for the standard (dual-asymmetric) spin $\mathbf{S}^\mathrm{standard}$ such a continuity equation has never been derived.

Is it possible to derive the properly conserved spin and orbital AM tensors in electromagnetic field theory from the Noether AM currents, without appealing to the dual symmetry? The key idea of our approach is to modify the spin and orbital AM fluxes in the canonical tensors $S^{\alpha\beta\gamma}$ and $L^{\alpha\beta\gamma}$, such that new tensors $\tilde{S}^{\alpha\beta\gamma}$ and $\tilde{L}^{\alpha\beta\gamma}$ properly satisfy the continuity equations [cf. the modification of the operators (2.13)]:

$$\boxed{\tilde{S}^{\alpha\beta\gamma} = S^{\alpha\beta\gamma} - \Delta^{\alpha\beta\gamma}, \quad \tilde{L}^{\alpha\beta\gamma} = L^{\alpha\beta\gamma} + \Delta^{\alpha\beta\gamma}, \quad \partial_\gamma \tilde{S}^{\alpha\beta\gamma} = \partial_\gamma \tilde{L}^{\alpha\beta\gamma} = 0.} \tag{3.11}$$

In doing so, we require that the spin-orbit correction $\Delta^{\alpha\beta\gamma}$ does not affect the spin and orbital AM densities $\mathbf{S}$ and $\mathbf{L}$ in Eqs. (2.7) or (2.8), and that it is properly antisymmetric:

$$\Delta^{\alpha\beta 0} = 0, \quad \Delta^{\alpha\beta\gamma} = -\Delta^{\beta\alpha\gamma}. \tag{3.12}$$

Comparing Eqs. (3.11) with (3.4), we find that the right-hand side of the canonical spin-continuity equation (3.4) needs to be represented as the total divergence of $\Delta^{\alpha\beta\gamma}$:

$$T^{\alpha\beta} - T^{\beta\alpha} = \partial_\gamma \Delta^{\alpha\beta\gamma}. \tag{3.13}$$

We implement the modification given by Eqs. (3.11)–(3.13) in the next subsections.

### 3.2. Explicit calculations for the standard electromagnetism.

In this subsection, we consider the standard electromagnetic field theory [11,12], which is based on the field Lagrangian $\mathcal{L} = -\frac{1}{4}F^{\alpha\beta}F_{\alpha\beta} = \frac{1}{2}(E^2 - B^2)$ being a functional of the gauge field (four-potential) $A^\alpha(r^\alpha)$. Because this Lagrangian is dual-asymmetric, the canonical dynamical characteristics and conservation laws also have dual-asymmetric form [12]. We recall that when performing calculations in covariant notations, we assume the Coulomb gauge (2.9), so the final results will not be Lorentz-covariant.

The canonical stress-energy tensor (3.1) and the corresponding orbital AM tensor (3.3) are

$$T^{\alpha\beta} = (\partial^\alpha A_\gamma)F^{\beta\gamma} - \frac{1}{4}g^{\alpha\beta}F^{\gamma\delta}F_{\gamma\delta}, \quad L^{\alpha\beta\gamma} = r^\alpha T^{\beta\gamma} - r^\beta T^{\alpha\gamma}. \tag{3.14}$$

The stress-energy tensor $T^{\alpha\beta}$ includes the canonical momentum density $\mathbf{P}^O = \mathbf{E}\cdot(\nabla)\mathbf{A}$, Eq. (2.8). The canonical spin tensor (3.3) and (3.4) reads

$$S^{\alpha\beta\gamma} = F^{\gamma\alpha}A^\beta - F^{\gamma\beta}A^\alpha. \tag{3.15}$$

Calculating the anti-symmetric part of the stress-energy tensor $T^{\alpha\beta}$ and *using the transversality of the vector-potential and electromagnetic field*, Eqs. (2.1) and (2.9), produces

$$T^{i0} - T^{0i} = \partial_k(E_k A_i), \quad T^{ij} - T^{ji} = \partial_k(\varepsilon_{ijl}B_k A_l). \tag{3.16}$$

By inspection, we can compare this result to the divergence term in Eq. (3.13) and determine the correction $\Delta^{\alpha\beta\gamma}$ to the spin and orbital AM tensors, Eqs. (3.11) and (3.12):

$$\Delta^{\alpha\beta 0} = \Delta^{00\gamma} = 0, \quad \Delta^{i0k} = -\Delta^{0ik} = E_k A_i, \quad \Delta^{ijk} = \varepsilon_{ijl}B_k A_l. \tag{3.17}$$

This is the key result of this paper, which yields the modified conserved spin and orbital AM currents $\tilde{S}^{\alpha\beta\gamma}$ and $\tilde{L}^{\alpha\beta\gamma}$, Eqs. (3.11).

The modified spin conservation law (3.11) with Eqs. (3.15) and (3.17) results in the continuity equation for the spin AM pseudo-vector $S_i = \frac{1}{2}\varepsilon_{ijk}\tilde{S}^{jk0} = \frac{1}{2}\varepsilon_{ijk}S^{jk0}$ and spin flux pseudo-tensor $\Sigma_{ij} = \frac{1}{2}\varepsilon_{ikl}\tilde{S}^{klj}$:

$$\partial_t S_i + \partial_j \Sigma_{ij} = 0,$$

$$S_i = (\mathbf{E}\times\mathbf{A})_i, \quad \Sigma_{ij} = \delta_{ij}(\mathbf{B}\cdot\mathbf{A}) - B_i A_j - B_j A_i. \tag{3.18}$$

This continuity equation for the spin AM is obtained here for the first time. It differs from the spin conservation suggested by Cameron *et al.* [23], Eq. (3.10), because Eq. (3.18) does not rely on the dual symmetry and is written for the standard spin AM density $\mathbf{S}^{standard}$. It is unrelated to the helicity pseudo-tensor or similarity with Lipkin's zilches, and is derived from the AM conservation in the form of the conserved rank-3 tensor $\tilde{S}^{\alpha\beta\gamma}$.

Equation (3.11) with Eqs. (3.14) and (3.17) result in the analogous conservation law for the pseudo-vector of the orbital AM, $L_i = \frac{1}{2}\varepsilon_{ijk}\tilde{L}^{jk0} = \frac{1}{2}\varepsilon_{ijk}L^{jk0}$ and its flux $\Lambda_{ij} = \frac{1}{2}\varepsilon_{ikl}\tilde{L}^{klj}$:

$$\partial_t L_i + \partial_j \Lambda_{ij} = 0,$$

$$L_i = [\mathbf{E}\cdot(\mathbf{r}\times\nabla)\mathbf{A}]_i, \quad \Lambda_{ij} = \varepsilon_{ikl}r_k\left[\varepsilon_{jmn}B_n(\partial_l A_m) + \frac{1}{2}\delta_{lj}(E^2 - B^2)\right] + B_j A_i. \tag{3.19}$$

To our knowledge, the orbital-AM conservation law is also derived here for the first time.

As it should be, the spin and orbital AM densities in Eqs. (3.18) and (3.19) coincide with the known **S** and **L** in Eqs. (2.7) or (2.8). At the same time, we will show in section 4 that the novel spin and orbital AM *fluxes* $\Sigma_{ij}$ and $\Lambda_{ij}$ in Eqs. (3.18) and (3.19) yield meaningful expressions for the spin and orbital AM in nonparaxial optical beams. We will see that these results are consistent with other approaches, but they correct the spin and orbital fluxes suggested previously in [20].

### *3.3. Results for the dual-symmetric electromagnetism.*

To restore the fundamental dual symmetry present in free-space Maxwell equations, but broken in the standard field Lagrangian and canonical Noether conservation laws, we recently suggested a dual-symmetric version of electromagnetic field theory [12] (see also [25]). The dual-symmetric electromagnetism is based on the Lagrangian $\mathcal{L} = -\frac{1}{8}\left(F^{\alpha\beta}F_{\alpha\beta} + G^{\alpha\beta}G_{\alpha\beta}\right)$ involving the second, dual gauge field (four-potential) $C^\alpha(r^\alpha)$ as $G^{\alpha\beta} = \partial^\alpha \wedge C^\beta$. Subject to constraint $G^{\alpha\beta} = *F^{\alpha\beta}$ (equivalent to Eqs. (2.2) and (2.3) in the Coulomb gauge), the dual-symmetric Lagrangian yields the same Maxwell equations of motion, but improved, dual-symmetric canonical conservation laws. It was shown in [12] that the dual-symmetric electromagnetism is a more consistent theory in free space than the standard one.

All equations in the dual-symmetric electromagnetism can be obtained from their standard-electromagnetism counterparts via symmetrization over the dual transformation (2.15). In particular, the stress-energy tensor (3.1), and the corresponding orbital AM tensor (3.3) become:

$$\boxed{T^{\alpha\beta} = \frac{1}{2}\left[\left(\partial^\alpha A_\gamma\right)F^{\beta\gamma} + \left(\partial^\alpha C_\gamma\right)*F^{\beta\gamma}\right], \quad L^{\alpha\beta\gamma} = r^\alpha T^{\beta\gamma} - r^\beta T^{\alpha\gamma}.} \quad (3.20)$$

This includes the canonical momentum density $\mathbf{P}^O = \frac{1}{2}\left[\mathbf{E}\cdot(\nabla)\mathbf{A} + \mathbf{B}\cdot(\nabla)\mathbf{C}\right]$, Eq. (2.16). In turn, the canonical spin tensor (3.3) and (3.4) reads

$$\boxed{S^{\alpha\beta\gamma} = \frac{1}{2}\left[F^{\gamma\alpha}A^\beta - F^{\gamma\beta}A^\alpha + *F^{\gamma\alpha}C^\beta - *F^{\gamma\beta}C^\alpha\right].} \quad (3.21)$$

Akin to Eq. (3.16), calculating the anti-symmetric part of the stress-energy tensor (3.20) and *using the transversality of the vector-potentials and fields* produces

$$T^{i0} - T^{0i} = \frac{1}{2}\partial_k\left(E_k A_i + B_k C_i\right), \quad T^{ij} - T^{ji} = \frac{1}{2}\partial_k\left[\varepsilon_{ijl}\left(B_k A_l - E_k C_l\right)\right]. \quad (3.22)$$

As before, we can write this part as the divergence term (3.13) and determine the correction $\Delta^{\alpha\beta\gamma}$ to the spin and orbital AM tensors, Eqs. (3.11) and (3.12):

$$\boxed{\Delta^{\alpha\beta 0} = \Delta^{00\gamma} = 0, \quad \Delta^{i0k} = -\Delta^{0ik} = \frac{1}{2}\left(E_k A_i + B_k C_i\right), \quad \Delta^{ijk} = \frac{1}{2}\varepsilon_{ijl}\left(B_k A_l - E_k C_l\right).} \quad (3.23)$$

This determines the modified conserved spin and orbital AM currents $\tilde{S}^{\alpha\beta\gamma}$ and $\tilde{L}^{\alpha\beta\gamma}$, Eqs. (3.11), in their dual-symmetric forms.

The modified spin conservation law (3.11), with Eqs. (3.21) and (3.23), results in the continuity equation for the spin AM pseudo-vector, $S_i = \frac{1}{2}\varepsilon_{ijk}\tilde{S}^{jk0} = \frac{1}{2}\varepsilon_{ijk}S^{jk0}$, and spin flux pseudo-tensor $\Sigma_{ij} = \frac{1}{2}\varepsilon_{ikl}\tilde{S}^{klj}$:

$$\partial_t S_i + \partial_j \Sigma_{ij} = 0,$$

$$S_i = \frac{1}{2}(\mathbf{E}\times\mathbf{A} + \mathbf{B}\times\mathbf{C})_i, \quad \Sigma_{ij} = \frac{1}{2}\left[\delta_{ij}(\mathbf{B}\cdot\mathbf{A} - \mathbf{E}\cdot\mathbf{C}) - B_i A_j - B_j A_i + E_i C_j + E_j C_i\right]. \quad (3.24)$$

This spin conservation law (3.24) coincides with the one suggested by Cameron *et al.* in [23] from the extension of the helicity conservation, Eqs. (3.8)–(3.10). Indeed, it is the dual-symmetric spin (2.16) that is equal to the flux of the helicity $H = \frac{1}{2}(\mathbf{B}\cdot\mathbf{A} - \mathbf{E}\cdot\mathbf{C})$ [12,23,25]. However, here the derivation of Eq. (3.24) relies solely on the AM conservation and the transversality conditions, and it is unrelated to the dual symmetry and helicity conservation.

Finally, from Eq. (3.11) with Eqs. (3.20) and (3.23), we obtain the continuity equation for the pseudo-vector of the orbital AM, $L_i = \frac{1}{2}\varepsilon_{ijk}\tilde{L}^{jk0} = \frac{1}{2}\varepsilon_{ijk}L^{jk0}$ and its flux $\Lambda_{ij} = \frac{1}{2}\varepsilon_{ikl}\tilde{L}^{klj}$:

$$\partial_t L_i + \partial_j \Lambda_{ij} = 0,$$

$$L_i = \frac{1}{2}\left[\mathbf{E}\cdot(\mathbf{r}\times\boldsymbol{\nabla})\mathbf{A} + \mathbf{B}\cdot(\mathbf{r}\times\boldsymbol{\nabla})\mathbf{C}\right]_i,$$

$$\Lambda_{ij} = \frac{1}{2}\left\{\varepsilon_{ikl}\varepsilon_{jmn}r_k\left[B_n(\partial_l A_m) - E_n(\partial_l C_m)\right] + B_j A_i - E_j C_i\right\}. \quad (3.25)$$

As it should be, the spin and orbital AM densities in Eqs. (3.24) and (3.25) coincide with the known dual-symmetric $\mathbf{S}$ and $\mathbf{L}$ in Eqs. (2.16) [12,15,23]. Interestingly, Maxwell equations allow conservations of spin and orbital AM in both dual-asymmetric and dual-symmetric forms, discussed in subsections 3.2 and 3.3. This means that the 'electric' and 'magnetic' parts of the spin and orbital AM densities are *separately-conserved quantities*, so that the dual symmetry is not essential here. This cannot be seen in the integral conservation laws (2.10), because the integral values of the 'electric' and 'magnetic' spin and orbital AM are equal to each other [15,30].

## 4. Monochromatic fields. Spin and orbital AM fluxes in optical beams.

### *4.1. Spin and orbital AM conservation in monochromatic fields.*

Here we consider applications of the above general results to the case of monochromatic fields (2.5), which are important in optics. Substituting Eqs. (2.5) and (2.6) in the spin and orbital AM conservation laws (3.18) and (3.19), we obtain time-averaged versions of these laws for the complex field amplitudes:

$$\partial_t \frac{1}{2\omega}\text{Im}(\mathbf{E}^*\times\mathbf{E})_i = -\nabla_j \frac{1}{2\omega}\text{Im}\left[\delta_{ij}\mathbf{B}^*\cdot\mathbf{E} - B_i^* E_j - B_j^* E_i\right] = 0, \quad (4.1)$$

$$\partial_t \frac{1}{2\omega}\text{Im}\left[\mathbf{E}^*\cdot(\mathbf{r}\times\boldsymbol{\nabla})\mathbf{E}\right]_i = -\nabla_j\left\{\frac{1}{2\omega}\text{Im}\left[\varepsilon_{jkl}B_l^*(\mathbf{r}\times\boldsymbol{\nabla})_i E_k + B_j^* E_i\right] + \frac{1}{4}\varepsilon_{ijk}r_k\left(|\mathbf{B}|^2 - |\mathbf{E}|^2\right)\right\} = 0. \quad (4.2)$$

Here the time derivatives obviously vanish because the complex field amplitudes (2.5) are time-independent. Nonetheless, the spin and orbital AM fluxes (under the gradient $\nabla_j$) represent meaningful physical characteristics of optical fields (see the next subsection).

The dual-symmetric versions of Eqs. (4.1) and (4.2) read

$$\partial_t \frac{1}{4\omega}\left[\text{Im}(\mathbf{E}^*\times\mathbf{E})_i + \text{Im}(\mathbf{B}^*\times\mathbf{B})_i\right] = -\nabla_j \frac{1}{2\omega}\text{Im}\left[\delta_{ij}\mathbf{B}^*\cdot\mathbf{E} - B_i^* E_j - B_j^* E_i\right] = 0, \quad (4.3)$$

$$\partial_t \frac{1}{4\omega} \text{Im} \left[ \mathbf{E}^* \cdot (\mathbf{r} \times \nabla) \mathbf{E} + \mathbf{B}^* \cdot (\mathbf{r} \times \nabla) \mathbf{B} \right]_i$$
$$= -\nabla_j \left\{ \frac{1}{4\omega} \text{Im} \left[ \varepsilon_{jkl} \left( B_l^* (\mathbf{r} \times \nabla)_i E_k + E_l (\mathbf{r} \times \nabla)_i B_k^* \right) + B_i^* E_j + B_j^* E_i \right] \right\} = 0 \; .$$
(4.4)

A spin continuity equation similar to Eq. (4.3) was suggested by Alexeyev *et al.* [19] and later considered by others [21,22]. However, in those papers, the authors considered *complex nonstationary fields*, i.e., complex solutions of real Maxwell equations (2.1). As far as we know, such fields do not exist in real world.

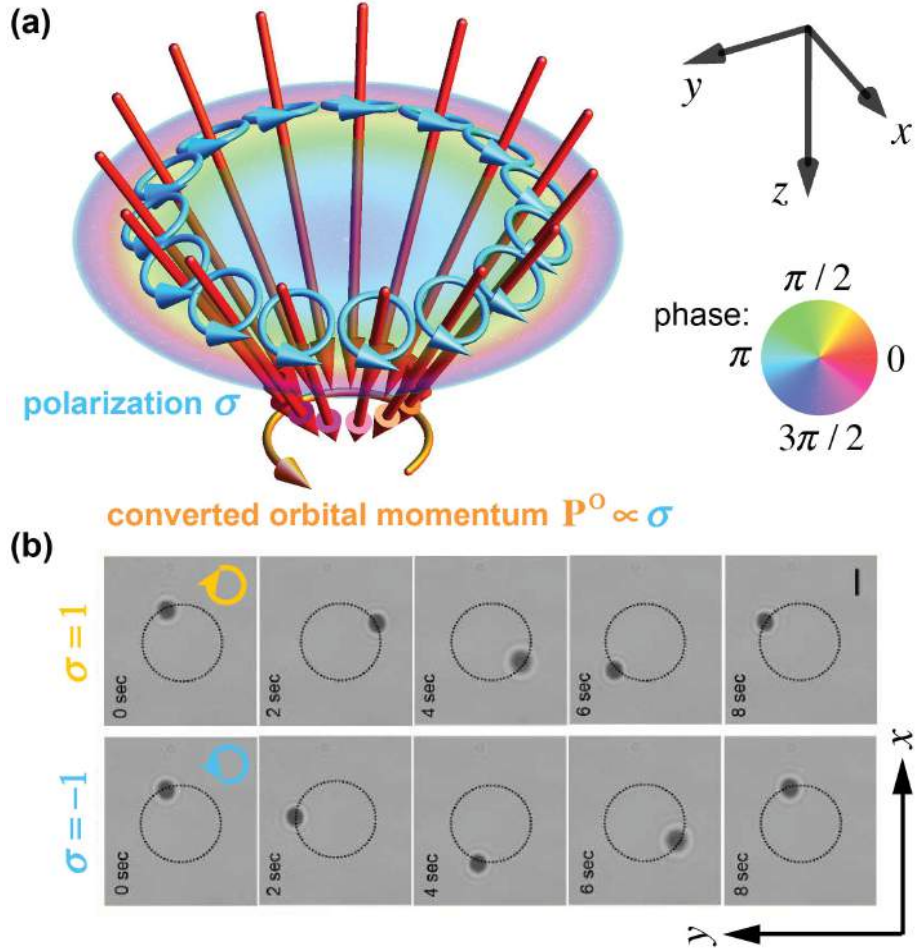

**Figure 2.** Spin-to-orbit AM conversion appears in nonparaxial optical fields in free space. A spherical geometry in momentum space (stemming from the electromagnetic wave transversality) and the Berry-phase contribution results in a *polarization*-dependent part of the *orbital* AM of a nonparaxial field [14,24]. For instance, **(a)** a tightly focused circularly-polarized ($\sigma = \pm 1$) optical beam *without* a vortex ($\ell = 0$) nonetheless exhibits a circulating orbital momentum $\mathbf{P}^O$ proportional to $\sigma(1 - \cos\theta_0)$ ($\theta_0$ being the characteristic focusing aperture angle). Experimental pictures **(b)** from [24c] demonstrate a *spin*-dependent *orbital* motion of a small particle in such a tightly focused field, i.e., the presence of the $\sigma$-dependent orbital AM. The spin-to-orbital converted part of the AM flux is precisely described by the $\Delta^{\alpha\beta\gamma}$ correction, Eqs. (3.11), (3.17), and (3.23), see subsection 4.2.

### *4.2. Spin and orbital AM fluxes in nonparaxial optical beams.*

As an application of the above general results, we consider the spin and orbital AM in nonparaxial optical vortex beams (e.g., Bessel beams) [1,13,14,20,36,37]. Straightforward classical-optics calculations and the separation of the integral spin and orbital AM in such beams faces some difficulties [36] because of the subtle 'surface AM' contribution [37]. One can efficiently calculate the spin and orbital AM using quantum-operator approaches [13,14], and there is a spin-dependent term in the orbital AM, which shows the *spin-to-orbit AM conversion* in nonparaxial fields [14] (see also [38]). This is an observable effect which appears upon the generation of a non-paraxial field: e.g., upon tight focusing or scattering of light [24] (see also and [18] for Dirac electrons), see Fig. 2. In 2002, Barnett [20] suggested to characterize the spin and orbital AM in nonparaxial beams via their *fluxes* integrated over the beam cross-section. This is a more natural approach (since the beams are delocalized states), but the fluxes suggested in [20] resulted in the perfect separation of the polarization-dependent spin and phase-dependent orbital AM parts *without* the spin-orbit effect.

Here we calculate the spin and orbital AM fluxes in a nonparaxial optical vortex beam, and show that the fluxes derived in our theory yield a result that is fully consistent with the quantum-operator approaches [13,14]. As we will see, our fluxes contain the spin-to-orbital conversion term due to the $\Delta^{\alpha\beta\gamma}$ correction, Eqs. (3.11), (3.17), and (3.23).

For the sake of simplicity, we consider a monochromatic field with a well-defined helicity, i.e., consisting of plane waves with the same circular polarization. Such fields are characterized by complex amplitudes $\mathbf{B} = -i\sigma\mathbf{E}$, where $\sigma = \pm 1$ is the helicity [14], and their characteristics are *equivalent* in the standard (electric-biased) and dual-symmetrized approaches. It is convenient to define the complex beam field as a 2D Fourier integral using spherical coordinates $(k,\theta,\phi)$ in the momentum $\mathbf{k}$-space (the $\theta = 0$ direction corresponding to the beam propagation) [14,37]:

$$\mathbf{E}(\mathbf{r}) \propto \int F(\mathbf{k})\mathbf{e}^\sigma(\mathbf{k})e^{i\mathbf{k}\cdot\mathbf{r}} d^2\mathbf{k}_\perp. \tag{4.5}$$

Here the complex scalar Fourier amplitude is $F(\mathbf{k}) = F(\theta,\phi) \propto f(\theta)e^{i\ell\phi}$ for the vortex beam with topological charge (orbital AM index) $\ell$, and $f(\theta) \propto \delta(\theta - \theta_0)$ for the Bessel beams. Next, $\mathbf{e}^\sigma(\mathbf{k}) = \mathbf{e}^\sigma(\theta,\phi)$ is the unit vector of the circular polarization (orthogonal to $\mathbf{k}$ due to the transversality). Finally, $\mathbf{k}\cdot\mathbf{r} = kz\cos\theta + k\rho\sin\theta\cos(\phi - \varphi)$, $(\rho,\varphi,z)$ are the cylindrical coordinates in the real space, and the integral (4.5) is taken over the transverse components of the wave vector: $d^2\mathbf{k}_\perp = k^2\sin\theta\cos\theta\, d\theta\, d\phi$ [14,37]. The unit polarization vectors are given by [14]:

$$\mathbf{e}^\sigma = \frac{\mathbf{e}_\theta + i\sigma\mathbf{e}_\phi}{\sqrt{2}}e^{i\sigma\phi}, \quad \mathbf{e}_\theta = (\cos\theta\cos\phi, \cos\theta\sin\phi, -\sin\theta), \quad \mathbf{e}_\phi = (-\sin\phi, \cos\phi, 0), \tag{4.6}$$

where $\mathbf{e}_\theta$ and $\mathbf{e}_\phi$ are written using their Cartesian components.

To characterize the spin and orbital AM in the $z$-propagating vortex beams, we calculate the spin and orbital AM *fluxes* through the transverse $(x,y)$ plane. These are given by

$$\langle \bar{\Sigma}_{zz} \rangle \propto \int \bar{\Sigma}_{zz}(\mathbf{r}) d^2\mathbf{r}_\perp, \quad \langle \bar{\Lambda}_{zz} \rangle \propto \int \bar{\Lambda}_{zz}(\mathbf{r}) d^2\mathbf{r}_\perp, \tag{4.7}$$

where the time-averaged fluxes $\bar{\Sigma}_{ij}$ and $\bar{\Lambda}_{ij}$ are the expressions under the gradient $\nabla_j$ in Eqs. (4.1) and (4.2), and $d^2\mathbf{r}_\perp = dxdy = \rho\, d\rho\, d\varphi$. Explicitly, from Eqs. (4.1), (4.2) and the helicity condition $\mathbf{B} = -i\sigma\mathbf{E}$, we find

$$\bar{\Sigma}_{zz} = \frac{\sigma}{2\omega}\left(|E_x|^2 + |E_y|^2 - |E_z|^2\right), \quad \bar{\Lambda}_{zz} = \frac{\sigma}{2\omega}\left[\text{Re}\left(E_y^* \partial_\varphi E_x - E_x^* \partial_\varphi E_y\right) + |E_z|^2\right]. \tag{4.8}$$

Here we took into account that $(\mathbf{r}\times\nabla)_z = \partial_\varphi$. The fluxes (4.8) differ from those suggested in [20] by the terms $-\sigma|E_z|^2/2\omega$ and $\sigma|E_z|^2/2\omega$ in the spin and orbital parts, respectively. These are corrections originating from the spin-orbit $\Delta^{\alpha\beta\gamma}$ correction, Eqs. (3.11), (3.17), and (3.23). These corrections in Eqs. (4.8) explicitly show the key role of the longitudinal $z$-component of the field (stemming from the transversality condition) in the spin-orbit interaction processes [24,39].

Since it makes sense to calculate the ratios of the spin and orbital AM to the energy, we also determine the energy flux in the beam. This is given by the $z$-component of the canonical momentum density (2.18) [2–5,12]:

$$\langle \overline{P}_z^{\,O} \rangle \propto \int \overline{P}_z^{\,O}(\mathbf{r}) d^2 \mathbf{r}_\perp, \quad \overline{P}_z^{\,O} = \frac{1}{2\omega} \text{Im}\left[\mathbf{E}^* \cdot \partial_z \mathbf{E}\right]. \tag{4.9}$$

Substituting now the beam field (4.5) into Eqs. (4.7)–(4.9), performing some vector algebra with Eqs. (4.6) and Fourier analysis of quadratic forms, we derive the integral energy, spin AM, and orbital AM fluxes in the beam:

$$\langle \overline{P}_z^{\,O} \rangle \propto \int |f(\theta)|^2 k \cos\theta \, d^2 \mathbf{k}_\perp, \tag{4.10}$$

$$\langle \overline{\Sigma}_{zz} \rangle \propto \sigma \int |f(\theta)|^2 \cos^2\theta \, d^2 \mathbf{k}_\perp, \tag{4.11}$$

$$\langle \overline{\Lambda}_{zz} \rangle \propto \int |f(\theta)|^2 \left[\ell + \sigma(1-\cos\theta)\right] \cos\theta \, d^2 \mathbf{k}_\perp. \tag{4.12}$$

Taking the simplest Bessel-beam case with $f(\theta) \propto \delta(\theta - \theta_0)$ [14], we obtain the finite ratios of the spin and orbital AM fluxes to the energy flux:

$$\boxed{\frac{\langle \overline{\Sigma}_{zz} \rangle}{\langle \overline{P}_z^{\,O} \rangle} = \frac{\sigma \cos\theta_0}{k}}, \quad \boxed{\frac{\langle \overline{\Lambda}_{zz} \rangle}{\langle \overline{P}_z^{\,O} \rangle} = \frac{\ell + \sigma(1-\cos\theta_0)}{k}}. \tag{4.13}$$

These results coincide with the ones for the integral spin and orbital AM values obtained in [14] using quantum-operator formalism (see also classical calculations in [38] and Dirac-electron calculations in [18]). The total (spin+orbital) AM (4.13) is equal to $(\sigma + \ell)/k$, but the separation is nontrivial because of the spin-to-orbital AM conversion term $\sigma(1-\cos\theta_0)/k$. This term originates from the $\Delta^{\alpha\beta\gamma}$ correction required for the local spin and orbital AM conservation laws, and it describes the observable effects of the spin-orbit interactions of light [14,24], Fig. 2.

This example demonstrates the validity of our general theory and its consistency with other approaches. In contrast, the fluxes suggested in [20] miss the spin-orbit term and effects, and, therefore, do not satisfy the conservation laws.

## 5. Conclusions

We have revisited the problem of the separation and description of the spin and orbital AM in free-space Maxwell fields. We have reviewed the previous approaches, both quantum and classical. Subtle but fundamental issues of the gauge invariance versus Lorentz covariance, and the presence of the dual symmetry/asymmetry have been discussed. We argued that the separation of the spin and orbital parts of the AM of light makes sense based on operational *local* measurements of these quantities, e.g., via probe particles (Fig. 1). In this manner, the gauge invariance of the spin and orbital AM densities is crucial, while the Lorentz covariance is broken by the probe: the quantities are characterized in a single laboratory reference frame.

The main remaining problem was the lack of local conservation laws (continuity equations) for the separated spin and orbital AM fluxes in electromagnetic field theory. Although the integral values of the spin and orbital AM are separately conserved quantities, their fluxes (following from the canonical AM tensor in field theory) do not satisfy the continuity equations [11,12]. We have resolved this problem in the present paper. Namely, we have found that the separation of the canonical AM flux into spin and orbital parts should be corrected with a spin-orbit term $\Delta^{\alpha\beta\gamma}$, which describes the observable spin-orbit interaction effects in nonparaxial fields (Fig. 2). In this manner, we have derived the modified spin and orbital AM tensors, which satisfy the local conservation laws and are consistent with previous quantum-operator approaches [13–17]. Our results correct the previous attempt to write the spin and orbital AM fluxes [20], which miss the spin-orbit terms. We also confirm the spin continuity equation suggested in [23] from the extended helicity conservation law. However, our spin and or orbital AM continuity equations are more general, because they do not involve the dual symmetry and can be written in the standard dual-asymmetric approach as well.

We have applied our theory to the case of nonparaxial optical vortex beams carrying both spin and orbital AM. Remarkably, the modified fluxes suggested in this work, precisely correspond to the integral spin and orbital AM values obtained earlier within the quantum-operator approach [13,14]. Thus, together with the previous works, our theory provides the complete and consistent description of the spin and orbital AM of free Maxwell fields in both quantum-mechanical and field-theory approaches.

Note that we have considered free-space fields, for which charges or currents should be considered as external entities perturbing the spin and orbital AM. The consideration of the changes in the spin and orbital AM induced by the presence of matter (charges, currents, or a continuous medium) is an important problem for future investigaton (see [5,8,13,19,21]). Finally, we note that the problem of the description of the spin and orbital AM is also highly important for quark and gluon fields in Quantum Chromodynamics in relation to the internal structure of the nucleon [40]. In this manner, the approaches presented in our work contain universal ideas that could be efficiently applied to other fields.

We acknowledge fruitful correspondence with A. Aiello, E. Leader, C. Lorcé, and F.W. Hehl. J.D. thanks Alexander Korotkov for the opportunity to complete this research. This work was partially supported by the RIKEN iTHES Project, JSPS-RFBR contract No. 12-02-92100, Grant-in-Aid for Scientific Research (S), ARO Grant No. W911NF-10-1-0334, and ARO MURI Grant No. W911NF-11-1-0268.

# References

1. *Optical Angular Momentum*, edited by L. Allen, S.M. Barnett, and M.J. Padgett (Taylor & Francis, London, 2003);
   *The Angular Momentum of Light*, edited by D.L. Andrews and M. Babiker (Cambridge University Press, Cambridge, 2012);
   L. Allen, M.J. Padgett, and M. Babiker, *Prog. Opt.* **39**, 291 (1999);
   S. Franke-Arnold, L. Allen, and M.J. Padgett, *Laser & Photon. Rev*. **2**, 299 (2008).
2. M.V. Berry, *J. Opt. A: Pure Appl. Opt.* **11**, 094001 (2009).
3. A. Bekshaev, K. Y. Bliokh, and M. Soskin, *J. Opt.* **13**, 053001 (2011).
4. K.Y. Bliokh, A.Y. Bekshaev, A.G. Kofman, and F. Nori, *New J. Phys.* **15**, 073022 (2013).
5. K.Y. Bliokh, A.Y. Bekshaev, and F. Nori, *Nat. Commun.* **5**, 3300 (2014).
6. A.T. O'Neil *et al.*, *Phys. Rev. Lett.* **88**, 053601 (2002);
   J.E. Curtis and D.G. Grier, *Phys. Rev. Lett.* **90**, 133901 (2003);
   V. Garcés-Chavéz *et al.*, *Phys. Rev. Lett.* **91**, 093602 (2003).
7. A. Canaguier-Durand *et al.*, *Phys. Rev. A* **88**, 033831 (2013).
8. K.Y. Bliokh, Y.S. Kivshar, and F. Nori, *Phys. Rev. Lett.* **113**, 033601 (2014).

9. V.B. Berestetskii, E.M. Lifshitz, and L.P. Pitaevskii, *Quantum Electrodynamics* (Pergamon, 1982);
   A.I. Akhiezer and V.B. Berestetskii, *Quantum Electrodynamics* (Interscience, 1965).
10. C. Cohen-Tannoudji, J. Dupont-Roc, and G. Grynberg, *Photons and Atoms* (Wiley, New York, 1989).
11. D.E. Soper, *Classical Field Theory* (Wiley, New York, 1976).
12. K.Y. Bliokh, A.Y. Bekshaev, and F. Nori, *New J. Phys.* **15**, 033026 (2013).
13. S.J. van Enk and G. Nienhuis, *J. Mod. Opt.* **41**, 963 (1994);
    S.J. van Enk and G. Nienhuis, *Europhys. Lett.* **25**, 497 (1994).
14. K.Y. Bliokh *et al.*, *Phys. Rev. A* **82**, 063825 (2010).
15. S.M. Barnett, *J. Mod. Opt.* **57**, 1339 (2010).
16. I. Bialynicki-Birula and Z. Bialynicki-Birula, *J. Opt.* **13**, 064014 (2011).
17. I. Fernandez-Corbaton *et al.*, arXiv:1308.1729v1.
18. K.Y. Bliokh, M.R. Dennis, and F. Nory, *Phys. Rev. Lett.* **107**, 174802 (2011).
19. C.N. Alexeyev, Y.A. Fridman, and A.N. Alexeyev, *Proc. SPIE* **4403**, 71 (2001);
    C.N. Alexeyev, Y.A. Fridman, and A.N. Alexeyev, *Ukr. J. Phys.* **46**, 43 (2001).
20. S.M. Barnett, J. Opt. B: Quantum Semiclass. Opt. **4**, S7 (2002).
21. J.E.S. Bergman *et al.*, arXiv:0803.2383v6.
22. M. Mazilu, *J. Opt. A: Pure Appl. Opt.* **11**, 094005 (2009).
23. R.P. Cameron, S.M. Barnett, and A.M. Yao, *New J. Phys.* **14**, 053050 (2012).
24. Y. Zhao *et al*., *Phys. Rev. Lett.* **99**, 073091 (2007);
    H. Adachi, S. Akahoshi, and K. Miyakawa, *Phys. Rev. A* **75**, 063409 (2007);
    Y. Zhao *et al*., *Opt. Express* **17**, 23316 (2009);
    C. Schwartz and A. Dogariu, *Opt. Express* **14**, 8425 (2006);
    Z. Bomzon and M. Gu, *Opt. Lett.* **32**, 3017 (2007);
    K.Y. Bliokh *et al*., *Opt. Express* **19**, 26132 (2011).
25. R.P. Cameron and S.M. Barnett, *New J. Phys.* **14**, 123019 (2012).
26. K.Y. Bliokh and F. Nori, *Phys. Rev. A* **86**, 033824 (2012);
    K.Y. Bliokh, Y.V. Izdebskaya, and F. Nori, *J. Opt.* **15**, 044003 (2013).
27. M.H.L. Pryce, *Proc. R. Soc. London A* **195**, 62 (1948).
28. I. Bialynicki-Birula and Z. Bialynicka-Birula, *Phys. Rev. D* **35**, 2383 (1987);
    B.-S. K. Skagerstam, arXiv:hep-th/9210054;
    A. Bérard and H. Mohrbach, *Phys. Lett. A* **352**, 190 (2006).
29. M.G. Calcin, *Am. J. Phys.* **33**, 958 (1965);
    D. Zwanziger, *Phys. Rev.* **176**, 1489 (1968);
    S. Deser and C. Teitelboim, *Phys. Rev. D* **13**, 1592 (1976).
30. In this context, we note that the inequalities in equations (2.47) and (3.36) in [12] should be equalities. The integral values of the dual-symmetric and dual-asymmetric spin or orbital AM are equal to each other, as it follows from the results of [15].
31. K. Y. Bliokh and F. Nori, *Phys. Rev. A* **85**, 061801(R) (2012);
    K.-Y. Kim *et al.*, *Phys. Rev. A* **86**, 063805 (2012).
32. F.J. Belinfante, *Physica* **6**, 887 (1939);
    F.J. Belinfante, *Physica* **7**, 449 (1940).
33. H.M. Wiseman, *New J. Phys.* **9**, 165 (2007);
    S. Kocsis *et al*., *Science* **332**, 1170 (2011).
34. S. Huard and C. Imbert, *Opt. Commun.* **24**, 185 (1978);
    S. Huard, *Can. J. Phys.* **57**, 612 (1979);
    S.M. Barnett and M.V. Berry, *J. Opt.* **15**, 125701 (2013).
35. D. Lipkin, *J. Math. Phys.* **5**, 696 (1964);
    D.J. Candlin, *Nuovo Cimento* **37**, 1390 (1965);
    T.W.B. Kibble, *J. Math. Phys.* **6**, 1022 (1965);
    R.F. O'Connel and D.R. Tompkins, *Nuovo Cimento* **39**, 391 (1965).

36. S.M. Barnett and L. Allen, *Opt. Commun.* **110**, 670 (1994).
37. M. Ornigotti and A. Aiello, *Opt. Express* **22**, 6586 (2014).
38. C.-F. Li, *Phys. Rev. A* **80**, 063814 (2009).
39. K.Y. Bliokh and Y.P. Bliokh, *Phys. Rev. Lett.* **96**, 073903 (2006);
    A.Y. Bekshaev, *Phys. Rev. A* **85**, 023842 (2012).
40. E. Leader and C. Lorcé, *Phys. Rep.* (2014, in press), doi: 10.1016/j.physrep.2014.02.010, arXiv:1309.4235;
    F.W. Hehl, arXiv:1402.0261.